\documentclass{iaus}
\usepackage{graphicx}

\title[IAUS267 ~~AGN evolution from large area and deep X--ray surveys]
{AGN evolution from large and deep X--ray surveys}

\author[Marcella Brusa]   %% give here short author list %%
{Marcella Brusa$^1$}

\affiliation{$^1$max Planck Institute f\"ur Extraterrestrische Physik, \\
Giessenbachstrasse, 1 DE-85748, garching bei M\"unchen, Germany \\
email: {\tt marcella@mpe.mpg.de}}

\pubyear{2009}
\volume{267}  %% insert here IAU Symposium No.
\pagerange{119--126}
\jname{Co-Evolution of Central Black Holes and Galaxies}
\editors{B.M.\ Peterson, R.S.\ Somerville, \& T.\ Storchi-Bergmann, eds.}
\begin{document}

\maketitle

\begin{abstract}
Over the last few years, the existence of mutual feedback effects between 
accreting supermassive black holes powering AGN and star formation in their 
host galaxies has become evident. This means that the formation and 
the evolution of AGN and galaxies should be considered as one and the 
same problem. As a consequence, the search for, and the characterization 
of the evolutive and physical properties of AGN over a large redshift 
interval is a key topic of present research in the field of observational 
cosmology. Significant advances have been obtained in the last ten years 
thanks to the sizable number of XMM-Newton and Chandra surveys, complemented 
by multiwavelength follow-up programs. I will present some of the recent 
results and the ongoing efforts (mostly from the COSMOS and CDFS surveys) 
aimed at obtaining a complete census of accreting Black Holes in the 
Universe, and a characterization of the host galaxies properties.
\end{abstract}


Since the launch in 1999 of both the XMM--{\it Newton} and {\it Chandra} satellites,
a large ($>$ 30) number of surveys covering a wide fraction of the area vs. depth plane
(see Fig.~1 in \cite{bh05} see also \cite{ryan09}) have been performed
and our understanding of AGN properties and evolution has received a major boost.
Thanks to vigorous programs of multiwavelength follow-up campaigns
which, since a few years, have became customary, sensitive X--ray observations turned
out to be highly efficient in unveiling  weak and/or elusive accreting black holes, in a
variety of otherwise {\it non-active} galaxies, such as (among others): X--ray Bright Optically
Normal Galaxies (XBONG, \cite{xbong}), 
Extremely Red Objects (e.g. \cite{b05}), 
Sub Millimeter Galaxies (e.g. \cite{alex05})
high-z starforming systems (e.g. \cite{daddi07}, \cite{f08}).
In many of these cases, the AGN responsible for the X-ray emission is overwhelmed at longer wavelength
by the host galaxy light and/or the obscuration might be connected to processes within the host galaxy itself, 
such as the star formation rate and the presence of dust lanes or starburst disks 
(see e.g. \cite{h09}). 
This suggests that the accretion activity (especially
in high-redshift sources) is unambiguously revealed thanks to the presence of a strong X--ray
emission (see e.g. discussion in \cite{cdfsmbff}) and therefore, the combination of both
X-ray and optical classifications can be crucial to fully assess the nature of the candidate AGN.

The high level of completeness in redshift determination for a large number
of X--ray selected AGN (up to a few thousands)  
has made possible a robust determination
of the luminosity function  and evolution of unobscured and mildly obscured AGN
which turned out to be luminosity dependent:
the space density of bright QSOs ($L_X > 10^{44}$ erg s$^{-1}$) peaks at z$\sim$ 2--3,
to be compared with the z$\sim$0.7--1 peak of lower luminosity  Seyfert galaxies
(\cite{ueda}, \cite{lafranca}, \cite{silverman}, \cite{ebrero}). 
Based on these works, \cite{marconi} and \cite{merloni} were 
the first to propose that SMBH undergo a ``anti--hierarchical" evolution, in the form a differential 
growth (earlier and faster for larger black holes).
This anti--hierarchical behavior observed in AGN evolution (similar to that observed in
normal galaxies, e.g. \cite{cowie}) provided an important, independent, confirmation that
the formation and evolution of SMBH and their host galaxies are likely different aspects
of the same astrophysical problem. 
In particular, the accretion onto supermassive Black Holes triggered by
galaxies mergers and/or collisions may provide the "feedback" needed in almost all recent models 
for galaxies formation and evolution, in order to recover the properties (masses,
luminoisity, clusetring) we observe today in local systems. 

In this general framework the differences between ``obscured"
and ``unobscured" AGN are no longer and uniquely described under
a zeroth-order geometrical unification model (in which they are simply related to
orientation effects, \cite{anto}), but can be interpreted as
due to the fact that the {\it same} object is observed in {\it different
evolutive} phases.
This hypothesis is further supported by the finding that absorption
is much more common at low luminosities (\cite{ueda}, \cite{maiolino}) and at high 
redshift (\cite{lafranca}, \cite{treister}, \cite{hasinger})
as emerges mainly from X--ray surveys.   
The luminosity and redshift dependence of the obscuring fraction may  
be naturally linked to the AGN radiative power (related to the intrinsic 
X--ray luminosity) which is able to ionize and expel gas and dust (more 
common at high-z) from the nuclear regions, nicely fitting the current 
framework of AGN formation and evolution sketched above (see, for examle,
\cite{h06} and this volume).

The complete picture is likely to be much more complicated, depending
on many other parameters (such as, e.g., the BH mass, the Eddington ratio, 
the QSO duty cycle), and in particular related to the complex ``lightcurves" 
of AGN, which, in turns, depend on the detailed hydrodynamics 
adopted in the simulations. 
A correct and complete identification of unobscured, obscured and highly
obscured AGN at all redshifts (and especially in the z=1-3 interval, where
most of the feedback is expected to happen) is therefore crucial
for a comprehensive understanding of the still little explored phase of
the common growth of SMBHs and their host galaxies. 

Here I highlight recent results the characterization of the high luminosity
tail of the X--ray luminosity function from the COSMOS survey,
and on the host galaxy properties of obscured AGN at cosmological distance, 
from ananalysis of X--ray selected sources in the Chandra Deep Field South
(CDFS). I also present evidence for an evolution of the M$_{BH}$-M$_{Bulge}$
relation as derived from optically selected BL AGN from the zCOSMOS survey
(almost all of them also X--ray emitters).

\section{The high-luminosity tail of X--ray selected AGN: results from the COSMOS survey}

The high completeness in optical identifications of the XMM-COSMOS sample 
(\cite{qsoz3}) 
and the availability of spectroscopic and photometric redshifts (\cite{mara})
allows us to estimate the number densities of AGNs as a function of luminosity and redshift. 
In calculating the binned number density evolution 
we considered only the subsample of sources detected in the hard (2-10 keV) 
band, an included in the flux limited sample (925 sources at $S_{\rm x}=6\times 10^{-15}$ erg 
s$^{-1}$ cm$^{-2}$; $\sim 60$\% with spectroscopic redshifts available). 
Since the hardness ratios between the hard and soft bands give the estimates of the absorption column density,
we present the XLF of the de-absorbed 2-10 keV luminosity. In our first analysis, we use the traditional 
$\sum 1/V_{\rm a}$ estimator (\cite{avni}).
Additional details on the method used in deriving the XLF are given in Brusa et al. (2010) 
and Miyaji et al. (in preparation). 

%%%%%%%%%%%%%%%%%%%%%%%%%%%%%%%%%%%%%%
\begin{figure}[b]
% \vspace*{-2.0 cm}
\begin{center}
 \includegraphics[width=3.4in]{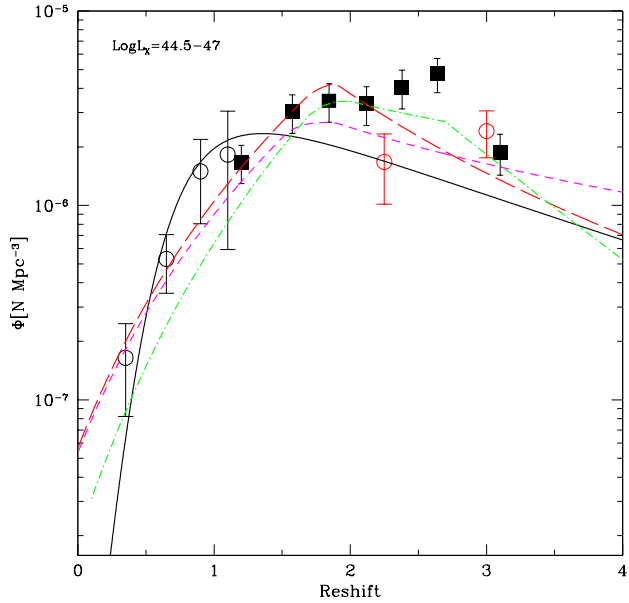}
% \vspace*{-1.0 cm}
 \caption{The number density evolution from the XMM-COSMOS 53 field AGN 
(squares) compared with recent results from Aird et al. (2010; 
circles, and black solid line),
Ebrero et al. (2009; short-dashed curve), Silverman et al. (2008; 
long-dashed curve), and the expectation of XRB synthesis 
models taken from \cite[Gilli et al. (2007, dot-dashed curve)]{gilli}.}
\label{fig:xlf_comp}
\end{center}
\end{figure}
%%%%%%%%%%%%%%%%%%%%%%%%%%%%%%%%%%%%%%

We compared our first estimates of the number density evolution of
the most luminous AGN population with recent measurements. 
The results are presented in Figure~\ref{fig:xlf_comp}, where the number 
density of XMM-COSMOS sources is plotted versus the redshift in the 
luminosity bin LogL$_{X}=44.5-47$ (squares).  
We only included the bins where the lowest $L_{\rm x}$ lowest $z$ 
boundary corresponds to a 2-10 keV flux above our limit 
for plotting. In the same figure we plot the data
points obtained from the AEGIS survey by \cite{aird}; circles), 
along with their best fit luminosity function (black curve).
The results based on 6 datapoints within z=1-3, with $\sim 20$ objects each 
may provide robust constraints on the shape of the XLF. 
In particular, the XMM-COSMOS data points seem to favour a higher
redshift (z$\sim2$) peak for the space density of luminous quasars 
as predicted by a LDDE parameterization, represented in the figure 
by the \cite[Ebrero et al. (2009,short-dashed curve)]{ebrero} 
and the \cite[Silverman et al. (2008, long-dashed curve)]{silverman} 
luminosity functions, though 
different in the details, and reproduced by XRB synthesis models 
(\cite{gilli}; dot-dashed curve) rather than 
the lower redshift (z$\sim 1$) peak expected from LADE model recently 
proposed by \cite[Aird et al. (2010)]{aird}.

\section{Black Hole and Star formation activity at z$>1$}

%%%%%%%%%%%%%%%%%%%%%%%%%%%%%%%%%%%%%%
\begin{figure}[b]
% \vspace*{-2.0 cm}
\begin{center}
 \includegraphics[width=3.4in]{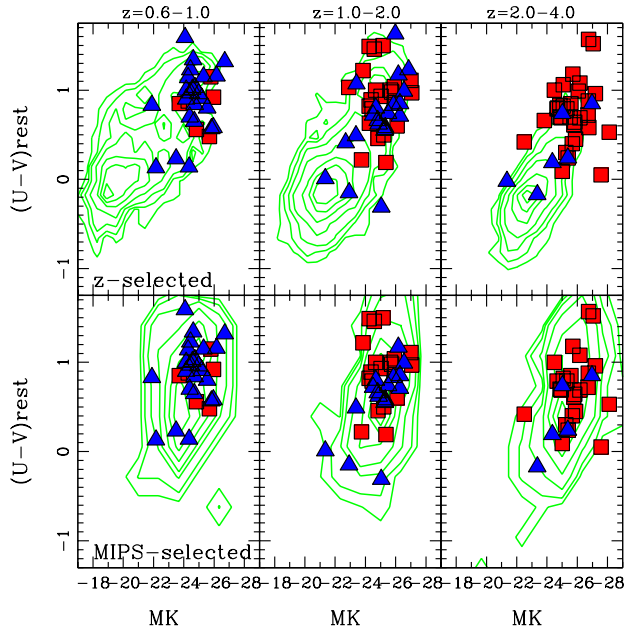} 
% \vspace*{-1.0 cm}
 \caption{U-V(rest frame) vs. the absolute K-band magnitude for the obscured AGN 
     (squares and triangles) and the underlying galaxy population (contours). The upper
     panels show the comparison of the X--ray selected sources with the
     optically-selected (z-band) galaxy population in three different redshift
     bins: z=0.6-1.0, z=1-2, and z=2.4 from left to right, respectively. The
     lower panels show the comparison for the subsample of objects detected
     also at 24 micron, in the same redshift bins.}
   \label{mkuv}
\end{center}
\end{figure}
%%%%%%%%%%%%%%%%%%%%%%%%%%%%%%%%%%%%%%
In order to study the host-galaxies of obscured AGN, 
we defined a sample of 116 ``bona fide'' obscured AGN, detected in
the 1Ms CDFS observation (\cite{alex03}) and for which
deep Infrared (IRAC and MIPS), K-band and multiwavelength optical 
photometry is available (from \cite{grazian}, the MUSIC survey)
We selected sources without broad lines in the optical spectra and with small 
optical nuclear emission with respect to the host galaxy optical emission. 

We found that the hosts of obscured AGN are redder in (U-V) rest frame 
than the overall galaxy population at the same redshift: in particular, 
obscured AGN mainly
populate the red sequence and the green valley in the color-magnitude
plots in agreement with the results of \cite{silv}.
For the MUSIC sample the U-V galaxy colors are
strongly correlated with the K band absolute magnitude
(Figure~\ref{mkuv}), and therefore
with the galaxy stellar mass, with the most massive systems having a
redder color.  The hosts of the obscured
AGN are therefore found in the red-massive tail of the distribution of 
optically selected galaxies in all three redshift bins considered
(see labels in the figure).  
AGN feedback is often invoked as one
of the main responsible for the observed galaxy colors (e.g. 
Nandra et al. 2007, \cite{hasinger}). 
However, it is well known that the main ingredient for nuclear
activity is the presence of a SMBH in galaxy
nuclei, and that SMBHs are found nearly exclusively in
massive galaxies (e.g. \cite{magorrian}).
Therefore, it is not truly surprising to find AGN hosted 
in massive galaxies and the simple presence of AGN in massive red galaxies 
is not enough to argue for a significant feedback effect on the 
observed colors, because of the strong color-mass correlation. 
Were AGN feedback responsible for the observed red colors, since
galaxy colors are strongly correlated with the galaxy mass and AGN are
found preferably in massive galaxies, then AGN feedback
should be also considered as one of the main players in the building 
of the galaxy mass-color correlation.

%%%%%%%%%%%%%%%%%%%%%%%%%%%%%%%%%%%%%%%%%%%%%%%%%%
\begin{figure}[b]
% \vspace*{-2.0 cm}
\begin{center}
 \includegraphics[width=3.4in]{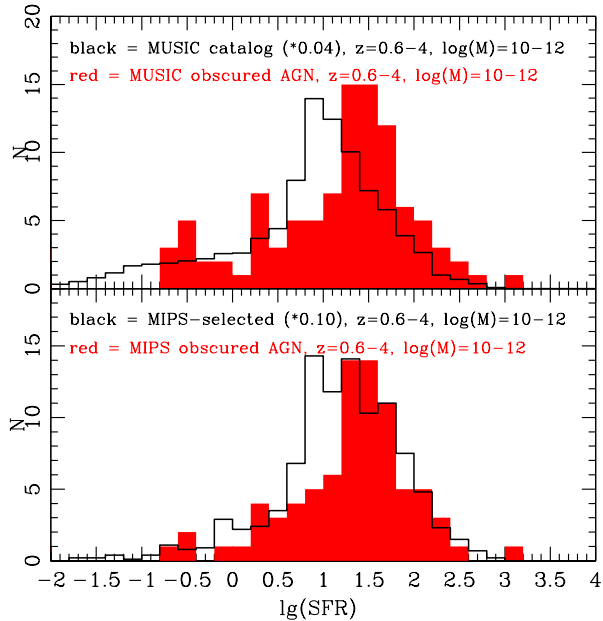} 
% \vspace*{-1.0 cm}
 \caption{The distribution of the SFR
for both the optically (upper) and MIPS (lower) selected sample of 
field galaxies  (open histogram) and obscured AGN (filled histogram) 
in the  redshift interval z=1-4 and in a specific stellar mass range 
(M$_*=10^{10}-10^{12}$  M$_\odot$).} 
   \label{sfr}
\end{center}
\end{figure}
%%%%%%%%%%%%%%%%%%%%%%%%%%%%%%%%%%%%%%%%%%%%%%%%%%

%%%%%%%%%%%%%%%%%%%%%%%%%%%%%%%%%%%%%%%%%%%%%%%%%%
\begin{figure}[b]
% \vspace*{-2.0 cm}
\begin{center}
 \includegraphics[width=3.4in]{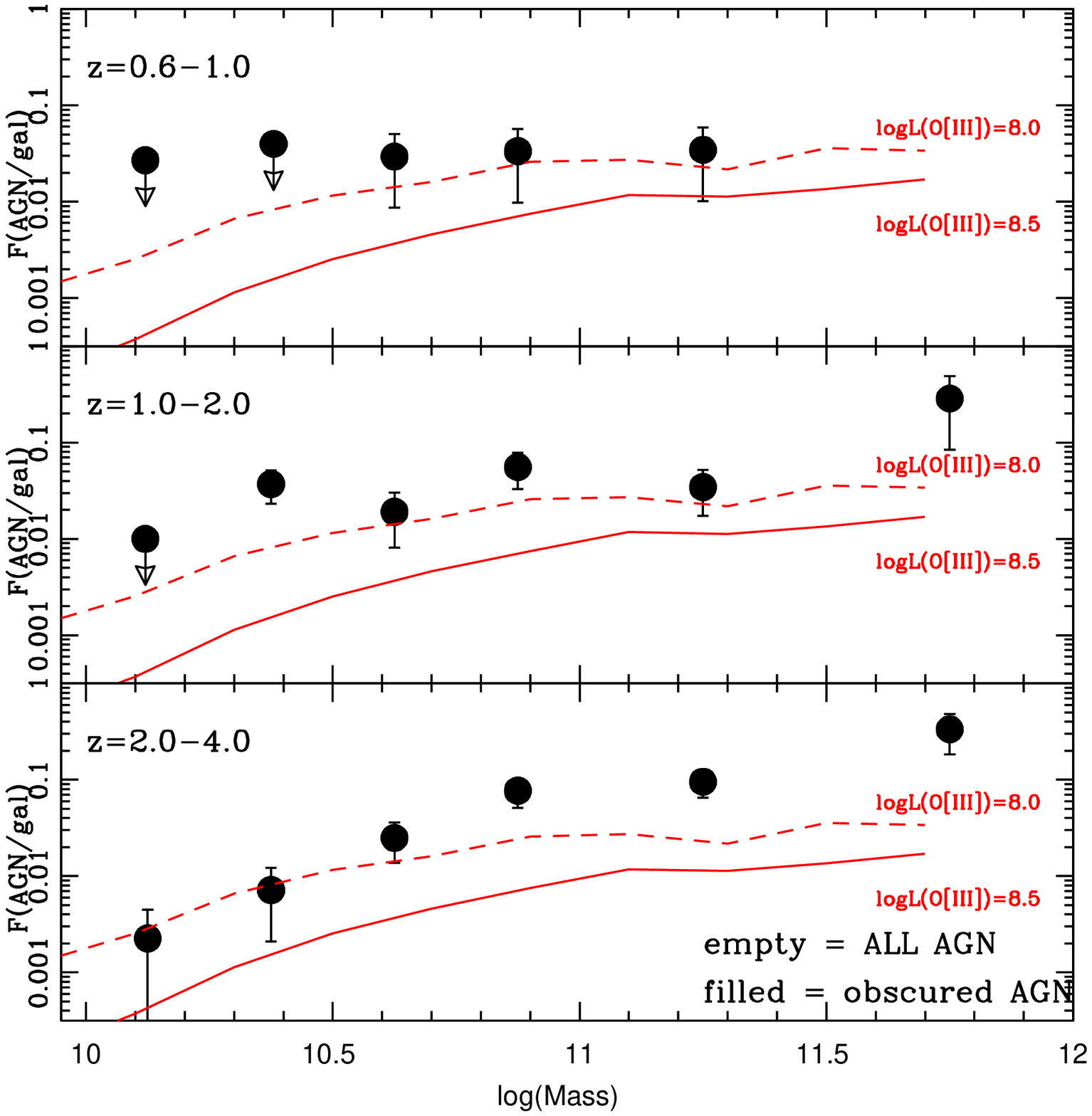} 
% \vspace*{-1.0 cm}
 \caption{Fraction of obscured AGN with L$_{X}>10^{43}$
erg s$^{-1}$ as a function
of the stellar mass in three different redshift bins.
Filled circles refer to the ``obscured'' AGN sample.
The dashed (continuous) lines represent the fraction of AGN with L(OIII)$>10^{8.0} (10^{8.5})$ L$_\odot$
in an optically selected sample in the local Universe (z$<0.2$) from the SDSS
(\cite{best}).} 
   \label{fract}
\end{center}
\end{figure}
%%%%%%%%%%%%%%%%%%%%%%%%%%%%%%%%%%%%%%%%%%%%%%%%%%

We found that about 2/3 of the obscured AGN
hosts at all redshifts show substantial ($>10$ M$_\odot$ yr$^{-1}$)
star formation activity (Figure~\ref{sfr}, where two histograms
of the SFR derivd from SED fitting are reported, for optical
and Infrared selected smaples) and about half 
live in galaxies which are still actively forming 
stars with respect to their mass.
For these sources, the observed red colors are
likely due to dust extinction rather than evolved stellar population. 
We then conclude that a significant fraction of obscured AGN 
live in massive, dusty star-forming galaxies with red optical colors. 

We compared the number of obscured AGN and of all X--ray selected AGN
to the number of field galaxies in broad bins of galaxy stellar mass
(M$_*=10^{10}-10^{12}$ M$_\odot$) and redshifts (z=0.6-1, z=1-2,
z=2-4). We find that the AGN fraction increases with the host
galaxy stellar mass, from $\sim$1\% at M$_*\sim 10^{10}$ M$_\odot$ to
$\sim$30\% at M$_*\sim 3\times10^{11}$ M$_\odot$ (see also \cite{yamada}),
and the actual trend 
of increasing AGN fraction  as a function of the stellar mass is probably 
steeper given the uncompleteness of the MUSIC sample at M$_{*}<10^{11}$ 
M$\odot$ (see Figure~\ref{fract}). 
We compared this trend with
that observed in the local Universe (\cite{best}) for AGN with 
luminosity above similar thresholds. While the observed trend is the same,
in all the investigated redshift bins the AGN fraction is 
higher than that observed in the local Universe, and it could likely be 
even higher.  
In fact, we are comparing here AGN selected with two different methods: forbidden
emission line luminosity (SDSS) and X-ray emission (GOODS). The latter
sample does not contain most Compton thick AGN.  
On the other hand, Compton thick AGN may well be present in [OIII] selected
AGN samples. 
The fraction of Compton thick AGN not directly
detected in deep Chandra surveys is estimated between 40\% and 100\%
of the X-ray selected AGN, using infrared selection or other
techniques (see \cite{donley}, \cite{f08} and references therein).
Therefore, under the simplest assumption that this Compton Thick AGN fraction is constant
with the galaxy mass, the discrepancy observed in Figure 10 can increase by up
of a factor of 2. \\
The fraction of active galaxies to the total galaxy population is
proportional to the AGN duty cycle. Our results would thus suggest
higher AGN duty cycles at z=2-4 than at z=0, in agreement with 
expectations from most recent semi-analytic models (e.g. \cite{menci}), 
in which at higher redshift the AGN activity is present in a 
larger number of galaxies than locally.  \\

The fact that the most luminous obscured AGN are found in the most massive galaxies
at all investigated redshifts may suggest that the L/L$_{Edd}$ of the obscured AGN 
is similar, particularly in the case of the most luminous sources (logLx$>43$ erg s$^{-1}$),
for which the threshold in luminosity introduces a bias against the sources 
accreting at lower rates in the lowest redshift bin. Assuming the local Magorrian 
relation between  M$_{BH}$ and M$_*$ (e.g \cite{mh03}) and a bolometric 
correction of 20 (e.g. \cite[Marconi et al. 2004]{marconi}) the median 
observed values of L$_X$/M$_*$ correspond to L/L$_{Edd}\sim0.1$. 
Although suffering from large uncertainties associated with the stellar mass and BH mass 
estimates, this value can be considered as representative of the accretion state of the most luminous, 
obscured AGN in the present sample.
Similar results are obtained for Chandra Deep Field North X--ray sources at z=2-4 
(\cite{yamada}) 
and are also typical of unobscured Type 1 AGN at z$>1$ (see \cite[Merloni et al. 2009]{merloni09}).

\section{The evolution of M$_{BH}$-M$_{Bulge}$ relation}

\begin{figure}[b]
% \vspace*{-2.0 cm}
\begin{center}
 \includegraphics[width=3.4in]{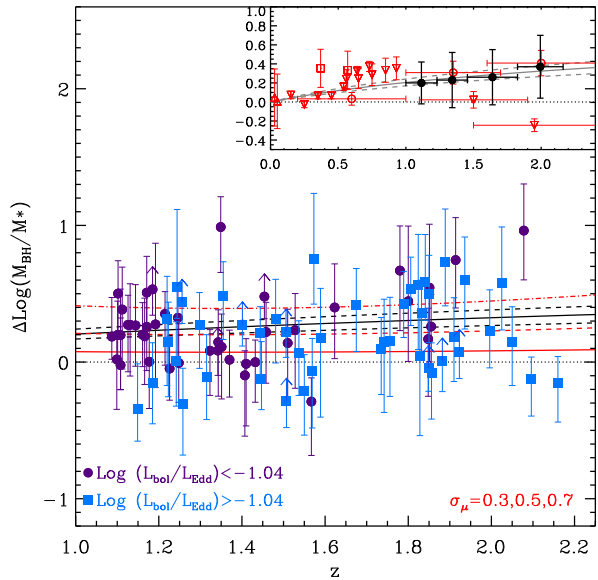} 
% \vspace*{-1.0 cm}
 \caption{Redshift evolution of the offset measured for zCOSMOS type 1 AGN form the local M$_{BH}$-M$_{*}$ 
 relation. Different symbols identify different ranges of Eddington ratios with upwards 
 arrows representing upper limit on the host galaxy mass. The solid black line shows the best fit
 obtained assuming an evolution of the form $\Delta Log(M_{BH}/M_*)$(z)=$\delta$(1+z) with $\delta$=0.68$\pm0.12$.
 The solid, dashed and dot-dashed lines show the bias due to an intrinisc scatter in the scaling 
 relation of 0.3, 0.5 and 0.7dex, respectively. In the inset we show a comparison of our data 
 (black circles, binned) with data collected from the literature.} 
   \label{mbh}
\end{center}
\end{figure}

Local scaling relations have proved themselves unable to unambiguously determine the physical 
nature of the SMBH-galaxy coupling. A large number of theoretical models for the AGN-galaxy interaction 
have been proposed, all tuned to reproduce the z=0 observations.One obvious way out of this impasse is 
the study of the scaling relations' evolution, which had, until recently, been limited to a handful 
of objects in narrow redshift windows (\cite{peng}, \& this volume; \cite{woo} \& this volume, \cite{jahnke}
\& this volume).

Within the zCOSMOS collaboration (a massive spectroscopic campaigns with $\sim10.000$ spectra of galaxies
and AGN, \cite{lilly}), we have pioneered a new method to unveil the intrinsic physical properties 
of AGN hosts, using the unprecedented multi-wavelength coverage of the COSMOS field. In \cite{merloni09},
we have been able to measure rest frame K-band luminosities and total stellar masses, $M_*$, of the hosts of 89 
broad-line (un-obscured) AGN in the redshift range $1<z<2.2$, for which we measured the black hole mass,
$M_{BH}$, through the virial method (see e.g. \cite{peterson} \& this volume). 
This sample constitutes the largest high redshift sample so far for which 
reliable black hole and galaxy masses are available. We found that, as compared to the local value, the average 
black hole to host galaxy mass ratio appears to evolve positively with redshift, with a best fit evolution of 
the form $(1+z)^{0.68\pm0.12}$ (see Figure~\ref{mbh}). A thorough analysis of observational biases induced by 
intrinsic scatter in the scaling relations reinforces the conclusion that an evolution of the $M_{\rm BH}-M_*$ 
relation must ensue for actively growing black holes at early times: either its overall normalization, or its 
intrinsic scatter (or both) appear to increase with redshift. \\

\par\noindent
{\bf Acknowlegdements} \\
It is a pleasure to acknowledge the contribution of Takamitsu Miyaji,  Andrea Comastri, Fabrizio Fiore, Andrea Merloni \& Angela Bongiorno in obtaining the results I presented at the conference,
along with the efforts of many people from the XMM-COSMOS, zCOSMOS and CDFS/MUSIC collaborations.
Many thanks to the organizers of the meeting, and in particular Bradley Peterson, for
offering me the possibility to attend this great symposium.

\end{document}